\documentclass[10pt]{article}
\usepackage{graphicx}

\def\Title#1{\begin{center} {\Large #1 } \end{center}}
\def\Author#1{\begin{center}{ \sc #1} \end{center}}
\def\Address#1{\begin{center}{ \it #1} \end{center}}

\newcommand\pubblock{\rightline{\begin{tabular}{l} Proceedings of the Fifth Annual LHCP\\ \pubnumber\\
         \pubdate  \end{tabular}}}

\newenvironment{Abstract}{\begin{quotation} \begin{center} 
             \large ABSTRACT \end{center}\bigskip 
      \begin{center}\begin{large}}{\end{large}\end{center} \end{quotation}}

\newenvironment{Presented}{\begin{quotation} \begin{center} 
             PRESENTED AT\end{center}\bigskip 
      \begin{center}\begin{large}}{\end{large}\end{center} \end{quotation}}

\def\Acknowledgements{\bigskip  \bigskip \begin{center} \begin{large}
             \bf ACKNOWLEDGEMENTS \end{large}\end{center}}

\def\beq{\begin{equation}}
\def\eeq#1{\label{#1}\end{equation}}
\def\eeqn{\end{equation}}

\def\beqa{\begin{eqnarray}}
\def\eeqa#1{\label{#1}\end{eqnarray}}
\def\eeqan{\end{eqnarray}}

\let\bar=\overbar

\def\Dslash{\not{\hbox{\kern-4pt $D$}}}
\def\dslash{\not{\hbox{\kern-2pt $\del$}}}

\def\msb{{\bar{\ssstyle M \kern -1pt S}}}

\usepackage{color}

\newcommand\pubnumber{ ATL-TILECAL-PROC-2017-013 }

\newcommand\pubdate{\today}

\def\affiliation{
On behalf of the ATLAS Collaboration, \\
Institute of Particle and Nuclear Physics, \\
Charles University, Prague, Czech Republic}

\begin{document}

\large
\begin{titlepage}
\pubblock

\vfill
\Title{  Performance of the ATLAS hadronic Tile calorimeter }
\vfill

\Author{ Michaela Mlynarikova  }
\Address{\affiliation}
\vfill
\begin{Abstract}
The Tile Calorimeter (TileCal) of the ATLAS experiment at the LHC is the central hadronic calorimeter designed for reconstruction of hadrons, jets, tau-particles and missing transverse energy. TileCal is a scintillator-steel sampling calorimeter and it covers the region of $|\eta| < 1.7$. The scintillation light produced in the scintillator tiles is transmitted by wavelength shifting fibers to photomultiplier tubes (PMTs). The analog signals from the PMTs are amplified, shaped and digitized by sampling the signal every 25 ns. The TileCal frontend electronics reads out the signals produced by about 10000 channels measuring energies ranging from $\sim 30$~MeV to $\sim 2$~TeV. Each stage of the signal production from scintillation light to the signal reconstruction is monitored and calibrated.
The performance of the calorimeter has been studied in-situ employing cosmic ray muons and a large sample of proton-proton collisions acquired during the operations of the LHC. Prompt isolated muons of high momentum from electroweak bosons decays are employed to study the energy response of the calorimeter at the electromagnetic scale. The calorimeter response to hadronic particles is evaluated with a sample of isolated hadrons and the modelling of the response by the Monte Carlo simulation is discussed. The calorimeter timing calibration and resolutions are studied with jets.
Results on the calorimeter operation and performance are presented, including the calibration, stability, absolute energy scale, uniformity and time resolution. These results show that the TileCal performance is within the design requirements and has given essential contribution to reconstructed objects and physics results.

\end{Abstract}
\vfill

\begin{Presented}
The Fifth Annual Conference\\
 on Large Hadron Collider Physics \\
Shanghai Jiao Tong University, Shanghai, China\\ 
May 15-20, 2017
\end{Presented}
\vfill
\end{titlepage}
\def\thefootnote{\fnsymbol{footnote}}
\setcounter{footnote}{0}
%

\normalsize 


\section{Introduction}

ATLAS \cite{1} is a general-purpose detector designed to reconstruct events from colliding protons and/or lead ions at the Large Hadron Collider. The Tile Calorimeter (TileCal) is the central hadronic calorimeter of the ATLAS experiment, designed for reconstruction of hadrons, jets, tau-leptons and missing transverse energy. It is a sampling calorimeter made of alternating layers of steel plates and scintillating tiles, the former serving as an absorber and the latter as an active medium. The calorimeter is divided into three longitudinal sections; one central long barrel ($|\eta|<1.0$) and two extended barrels ($0.8 < |\eta| < 1.7$) on either side of the long barrel. Each of these barrels is segmented into 64 modules in the azimuthal plane, with readout cells being organised into three radial layers A, BC and D. For layer A and BC the cell segmentation in the $\eta - \phi$ plane is $\Delta \eta \times \Delta \phi = 0.1 \times 0.1$ and for the outermost layer D $\Delta \eta \times \Delta \phi = 0.2 \times 0.1$. The light generated in each scintillator is collected at two edges, and then transported to photomultiplier tubes (PMTs) by wavelength shifting fibers. The analog signal from each PMT is amplified using two gains (low and high gain, gain ratio 1:64) and shaped by a passive shaper. The shaped signals are sampled and digitized by a 10-bit analog-to-digital converters (ADCs) every 25~ns \cite{2}.  

\section{Energy reconstruction and calibration procedure}
The Optimal Filtering technique \cite{OF} is used for the reconstruction of the signal amplitude ($A$), time phase ($\tau$), and pedestal ($p$) for each channel using the ADC count of each sample $S_i$ taken at time $t_i$:

 \begin{equation} \nonumber
      A = \sum\limits_{i=1}^{7} a_i S_i, \qquad A \tau = \sum\limits_{i=1}^{7} b_i S_i, \qquad p = \sum\limits_{i=1}^{7} c_i S_i, 
    \end{equation}
where the weights ($a_i$, $b_i$, $c_i$) are derived using the reference pulse shape to minimize the resolution of the amplitude and time.

A set of calibration constants $C_i$ is then used to convert the reconstructed channel amplitude in ADC-counts to energy

\begin{equation}
      E~[\mathrm{GeV}] = A~[\mathrm{ADC}] \times C_\mathrm{laser} \times C_{\mathrm{Cs}} \times C_{\mathrm{CIS}}~[\mathrm{pC/ADC}] / C_{\mathrm{TB}}~[\mathrm{pC/GeV}].
      \label{eq:2}
\end{equation}
The constant $C_{\mathrm{TB}} = 1.05$~pC/GeV is the global electromagnetic scale of the TileCal determined in dedicated beam tests, during which $11\%$ of the production modules were tested with electron beams \cite{3}. The other three constants are obtained using dedicated calibration systems. Each of them controls different parts of the readout chain, as described in the following.

\begin{figure}[t]
\begin{minipage}[c]{0.49\textwidth}
\includegraphics[width=0.95\textwidth]{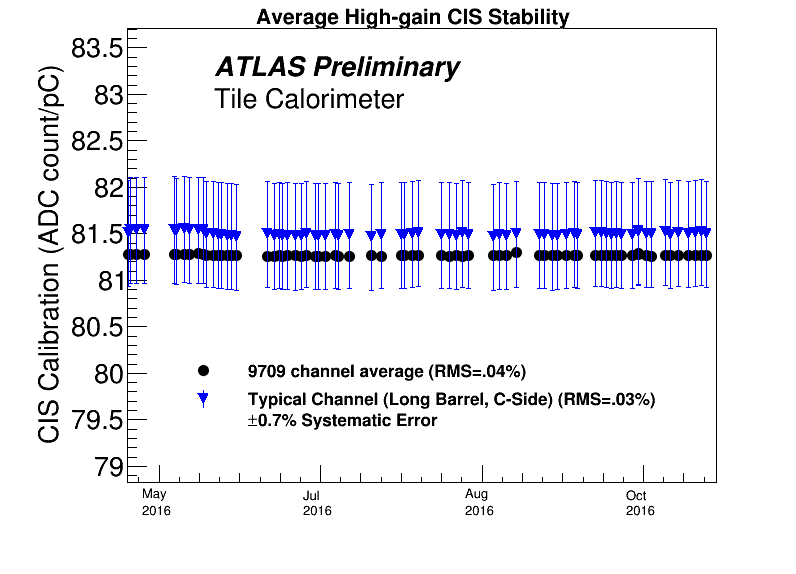}
\end{minipage}
\hspace{5mm}
\begin{minipage}[c]{0.49\textwidth}
\includegraphics[width=0.95\textwidth]{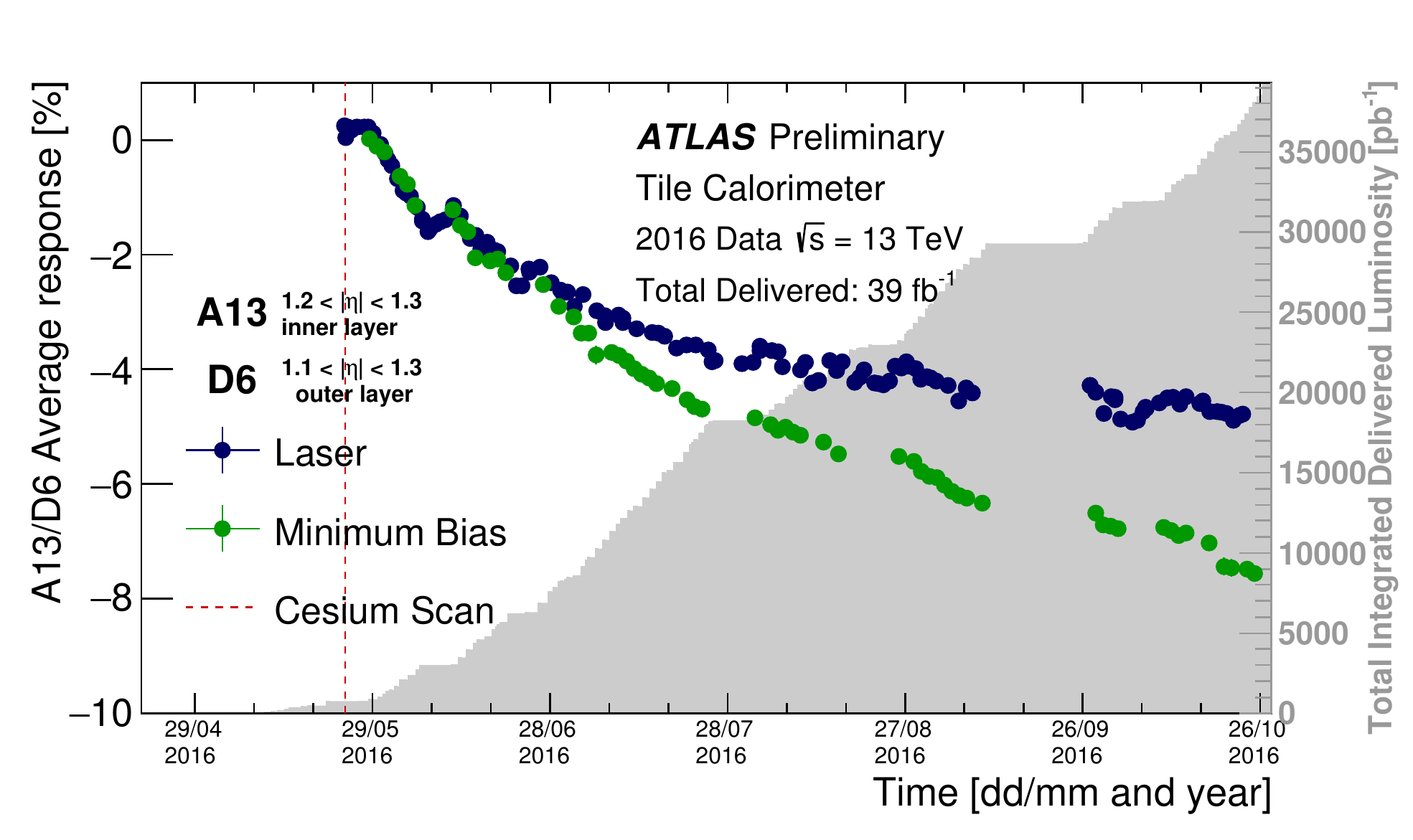}
\end{minipage}
\caption{Left: CIS calibration constant averages of all high-gain channels for each CIS calibration from May 2016 to Oct 2016.
Right: Variation of the response to Minimum Bias (green) and laser (blue) for A13 cells with respect to reference cell D6 as a function of days of data taking in 2016. The red line indicates the date of the last Cesium calibration. The integrated luminosity is shown with grey color in the background. Figures are taken from Ref.~\cite{Calib}. }
\label{fig:figure1}
\end{figure}

The conversion from ADC counts to pC, $C_{\mathrm{CIS}}$, is calculated for each channel using a charge injection system (CIS), which injects a known charge in both high gain and low gain readouts, and the response of the electronics is measured. The overall stability of the calibration factor is at the level of $0.02\%$ (Fig.~\ref{fig:figure1} left). This system is also used to monitor the front-end electronics and correct for non-linearities. The laser calibration system sends laser light pulses of adjustable intensity directly to each PMT and the main purpose of this system is to measure individual PMT gain variations. These variations are corrected with the factor $C_\mathrm{laser}$. Laser events are also used during the physics data taking to monitor the time calibration. The laser calibration is typically taken together with the CIS calibration twice per week. The Cesium calibration system is used to equalise the cell response. A movable $^{137}Cs$ $\gamma$-source emitting 0.662~MeV photons is hydraulically driven through special tubes across each TileCal cell, illuminating every single scintillator.

In addition to the aforementioned calibration systems, there is the Minimum Bias System which is used to monitor the instantaneous luminosity in ATLAS and which uses the same integrator readout as the Cesium calibration system. Both the Cesium and the Minimum Bias system measure the signal coming from scintillators and they are both expected to have the same variation in PMT response over time. The difference between the Laser and Minimum Bias/Cesium response allows to estimate the effect of the scintillator irradiation, which was found to be at the level of $3\%$ for the most irradiated cell A13 in 2016, see Fig.~\ref{fig:figure1} right. Cesium calibrations are taken approximately every month, however due to a water leak in TileCal at the beginning of 2016 operations, the Cesium calibration system was not available during the whole year 2016. 
To maintain the TileCal performance the laser calibration was taken every day in 2016 and the cell equalisation was checked using minimum bias events for all cells. 

All calibration systems together allow to maintain the electromagnetic scale across the calorimeter with precision of about $1\%$.

\section{Performance with single particles}
\begin{figure}[t]
\center
\hspace{5mm}
\begin{minipage}[]{0.4\textwidth}
\includegraphics[width=0.95\textwidth]{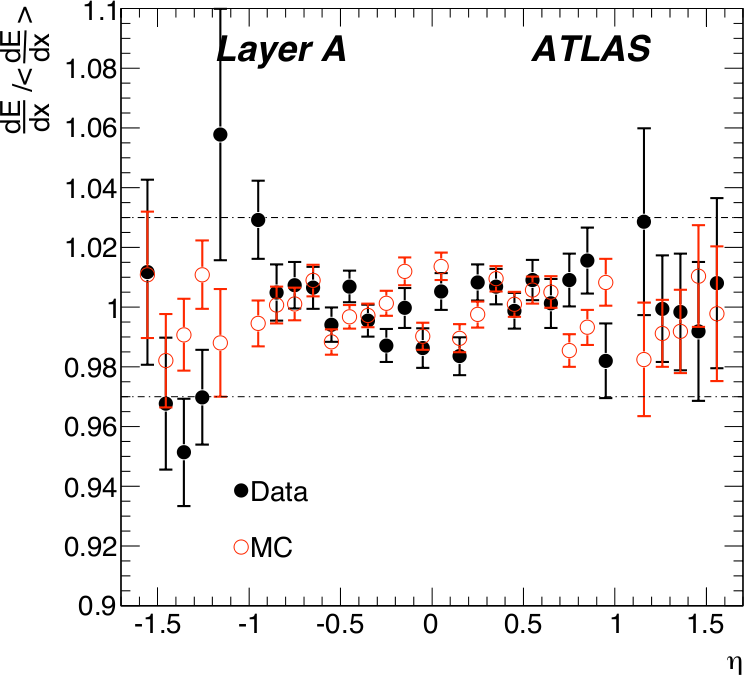}
\end{minipage}
\begin{minipage}[]{0.49\textwidth}
\includegraphics[width=0.95\textwidth]{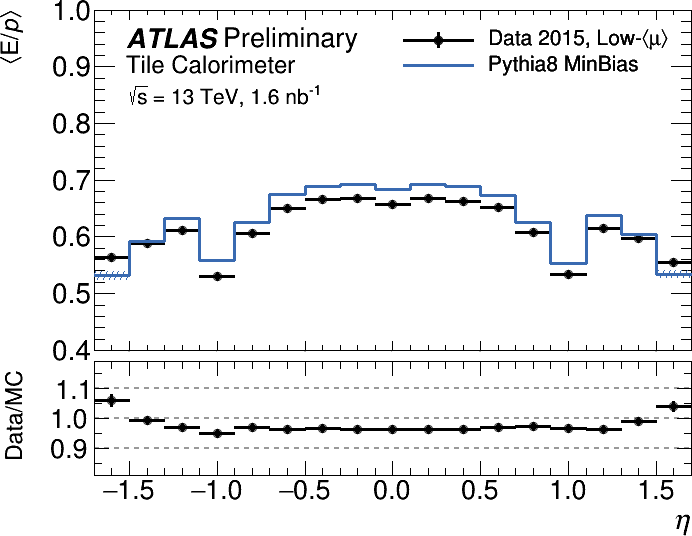}
\end{minipage}
\caption{Left: Uniformity of the cell response to cosmic muons expressed in terms of the normalised truncated mean of $dE/dx$, as a function of $\eta$ for radial layer A~\cite{cosmicmu}. Right: Energy over momentum $E/p$ as a function of $\eta$ for isolated hadron track using 2015 data~\cite{singleparticle}.}
\label{fig:figure2}
\end{figure}

The performance of the TileCal, i.e.~methods used to reconstruct and calibrate object and correct for problematic regions of the detector, was evaluated with single isolated particles. 
Single isolated muons were used to check the electromagnetic scale across the detector, using the ratio of the energy deposited to the path length traveled by the muon in a cell $dE/dx$.
For muon energies below 100~GeV this ratio is approximately constant.
The muons used either come from cosmic rays or from heavy gauge boson decays in proton-proton collisions. 
The uniformity of the cell response to cosmic muons \cite{4} expressed in terms of the normalised truncated mean of $dE/dx$, as a function of $\eta$ for radial layer A is shown in Fig.~\ref{fig:figure2} left. 
The response is integrated over all cells in each $\eta$ bin and the results for data and Monte Carlo are normalized to their averages.
The calorimeter cell response to collision muons from $W \rightarrow \mu \nu$ decays using 2012 experimental and simulated data can be found in Ref.~\cite{singleparticle}.

Due to non-uniformities in the individual tiles and due to the exponential attenuation of the light in the scintillators, the measured deposited energy in a cell (calculated as the sum of the two PMT energies) depends on the exact position of the local energy deposit. The transverse size of a hadronic shower is much bigger than the size of a single tile and therefore the effect is not visible, but for a single muon track it is possible to obtain the energy profile as a function of $\phi$ of the track. The values of $dE/dx$ obtained from collision events with muons produced in $W \rightarrow \mu \nu$ decays as a function of $\Delta \phi$ -- the difference in $\phi$ between the muon impact point and the cell center --- can be found in Ref.~\cite{Ushape}. Variations up to $10\%$ were observed with data collected until 2013 (run-1) and this profile is used now for simulations for run-2 (since 2015).

The performance of the TileCal was also studied with single isolated charged hadrons, using the ratio of the energy measured in TileCal, to  the momentum of their track reconstructed in the inner detector $E/p$. 
A signal compatible with a minimum ionizing particle in the electromagnetic calorimeter in front of the TileCal was required, to avoid any bias of the TileCal response due to interactions in upstream material.
The mean $E/p$ as a function of $\eta$ is shown in Fig.~\ref{fig:figure2} right for data taken in 2015 and Monte Carlo simulations.
Most of the data and MC agree within $5\%$.

\section{Conclusions}
Compared to previous years, TileCal has shown its best operation performance during the 2016 data taking with less than $1\%$ of non-operational cells by the end of collisions. Its calibration and monitoring systems guarantee stability in time of the calorimeter response within $1\%$. The energy measurement was validated in-situ showing that the TileCal performs within the design parameters, and as such has provided the foundation of many physics results that are based on the reconstructed objects.
 


\Acknowledgements
The support from Ministry of Education, Youth and Sports of the Czech Republic is acknowledged.


\begin{thebibliography}{99}


  \bibitem{1} ATLAS Collaboration, {\it The ATLAS Experiment at the CERN Large Hadron Collider, JINST} {\bf 3} (2008) S08003.
  
  \bibitem{2} K. Anderson et al, {\it Design of the fron-end analog electronics for the ATLAS Tile Calorimeter, NIM} A {\bf 551} (2005).
  
  \bibitem{OF} W. E. Cleland and E. G. Stern, NIM {\bf A338}, 467–97 (1994).
  
  \bibitem{3} P. Adragna et al, {\it Testbeam studies of production modules of the ATLAS Tile Calorimeter, NIM} A {\bf 606} (2009).

  \bibitem{Calib} https://twiki.cern.ch/twiki/bin/view/AtlasPublic/ApprovedPlotsTileCalibration.
  
  \bibitem{4} ATLAS Collaboration, {\it Readiness of the ATLAS Tile Calorimeter for LHC collisions},
Eur. Phys. J. C {\bf 70} (2010) 1193.
  
  \bibitem{singleparticle} https://twiki.cern.ch/twiki/bin/view/AtlasPublic/ApprovedPlotsTileSingleParticleResponse.
 
 \bibitem{Ushape} https://twiki.cern.ch/twiki/bin/view/AtlasPublic/ApprovedPlotsTileUShape.
 
 \bibitem{cosmicmu} https://twiki.cern.ch/twiki/bin/view/AtlasPublic/ApprovedPlotsTileEnergyCalibration.
 
    

  
  
  
  
  
  


\end{thebibliography}
\end{document}